# Mobile Re-Finding of Web Information Using a Voice Interface

Robert G. Capra
Department of Computer Science, Virginia Tech
660 McBryde Hall (0106)
Blacksburg, VA 24061 USA
1 540 231 6931

rcapra@vt.edu

Manuel A. Pérez-Quiñones
Department of Computer Science, Virginia Tech
660 McBryde Hall (0106)
Blacksburg, VA 24061 USA
1 540 231 2646

perez@vt.edu

## ABSTRACT
Mobile access to information is a considerable problem for many users, especially to information found on the Web. In this paper, we explore how a voice-controlled service, accessible by telephone, could support mobile users' needs for re-finding specific information previously found on the Web. We outline challenges in creating such a service and describe architectural and user interfaces issues discovered in an exploratory prototype we built called WebContext.

We also present the results of a study – motivated by our experience with WebContext – to explore what people remember about information that they are trying to re-find and how they express information re-finding requests in a collaborative conversation. As part of the study, we examine how end-user-created Web page annotations can be used to help support mobile information re-finding. We observed the use of URLs, page titles, and descriptions of page contents to help identify waypoints in the search process. Furthermore, we observed that the annotations were utilized extensively, indicating that explicitly added context by the user can play an important role in re-finding.

## Categories and Subject Descriptors
H.1.2 [Models and Principles]: User/Machine Systems—Human factors; Human information processing; H.3.5 [Information Storage and Retrieval]: Online Information—Web-Based Services; H.5.2 [Information Interfaces and Presentation]: User Interfaces—Natural language; Voice I/O

## General Terms
Design, Experimentation, Human Factors

## Keywords
Web information re-finding, personal information repositories, information management, voice user interfaces, speech recognition, conversation analysis, dialog processing

## 1. INTRODUCTION
Computer users today have many needs to store, re-find, and re-use electronic information, yet these tasks are neither well understood nor well supported by existing software tools and interfaces. People use the web to find information, but often have trouble organizing and re-finding information they have found [JBD01]. Re-finding information found on the Web can be especially difficult for mobile users with limited computing resources.

A considerable amount of research is conducted to help people *find* information on the web. For example, consider research on web search engines, work in the area of recommender systems and web personalization, and research on information foraging. However, re-finding information on the web is a considerably different process that has received "relatively less" [JBD01, p.119] investigation than the problem of how to find things in the first place. Results from a GVU study of web usage [GVU98] suggest that re-finding web pages is a problem and that users have trouble organizing information found on the web. Given that re-finding and organizing information found on the web is a problem for users, a better understanding of the re-finding process is needed to help guide the development of tools to assist users in their information re-finding tasks.

The research presented here focuses on investigating how voice interfaces can be used to support mobile users' web information re-finding needs. We present details of an initial prototype system developed to support remote access to re-finding of information first found on web pages. This system, called WebContext [CPR01], was developed to provide a testbed for us to explore this topic. We also present the results of a study conducted to examine the strategies people use for information re-finding. Understanding how people approach and converse about re-finding web information is an essential element of designing an effective voice interface to support these needs. As part of the study, pairs of human participants engaged in a set of collaborative, information re-finding tasks.

### 1.1 Mobile Information Re-Finding
We are particularly interested in understanding and supporting the information re-finding needs of mobile users. Mobile (remote) access to information is an important dimension of re-use [JBD01] and can be especially challenging for mobile users. A recent study found that mobile workers rely heavily on the use of cell phones to enlist the help of co-workers "back at the office" to retrieve information [POS+01]. Users would call back to the office to gain access to information and resources "by proxy" by engaging a co-worker to collaboratively work on the tasks over the telephone. This type of use could include re-finding, managing, and composing information.

In order to build effective tools to support re-finding, it is important to understand how people go about re-finding information. There are several components to this. First, users may organize information to facilitate re-use. How do they do this and what tools are effective to support organization for re-use? Second, what do people remember about the information they are trying to re-find? This may include information that appeared in the same document as the information being sought, or may be meta-information such as the time of day they originally saw the information, the particular computer they were using, or how they navigated to the information. Third, it is important to know how people go about the process of trying to re-find information. Obviously, this will depend a great deal on the specific tools and interface being used to accomplish the re-finding, but we can, in general, try to gain an understanding of the processes and approaches taken by users in trying to re-locate information that they have seen before. While we do not attempt to answer each of these research questions here, they are essential elements in the study of information re-finding and we do address aspects of them as they relate to our research.

In mobile situations, users are likely to be trying to re-find information using a mobile device (PDA, cell phone) that is different than the device on which the information was originally found (home or work computer). The cues and utilities of the desktop computer may not be present when the user is trying to re-access the information. We view these mobile interfaces as ways to provide *directed access to satisfy specific information needs rather than as replacements for desktop applications*. These directed interfaces need to support the artifacts and vocabulary that people use to communicate and reason about their re-finding needs in order to help bridge the gap between device capabilities and modalities. Thus, understanding the artifacts and processes employed by users is a significant goal of this research. This goal also stems from our interest in supporting multiple types of interfaces for mobile information access, including voice interfaces [CPR01].

## 1.2 Telephone Voice Interfaces

Voice user interfaces, despite their limitations, are proving to be an effective means to provide mobile access to information through the use of both wireline and wireless telephones. For example, services such as Wildfire [www.wildfire.com] provide voice access to voicemail and personal contact information. The emergence of VoiceXML as an implementation tool has spurred general use services such as TellMe [www.tellme.com] and BeVocal [www.bevocal.com] that provide toll-free access to news, weather, sports, and stock information using voice interfaces over the telephone. Given the expanding market for telephone-based voice services, a large number of users may access telephone-based VoiceXML browsers in the near future. Such access opens up a wide array of service options, including providing users with voice access to more personalized and personal information remotely via telephone.

Developing voice interfaces for re-finding requires an understanding of how users verbalize their information re-finding needs and strategies. We need to explore how people phrase information re-finding requests and to understand the structure of a collaborative conversation for information re-finding.

In this paper, we present results and observations from a controlled, laboratory study we conducted to gain a better understanding of how people approach information re-finding: what they remember (recall and recognize) when trying to re-find, and how context and contextual information is used in the re-finding process. The results presented are relevant to research in the area of web information re-finding and also to the area of personalized voice interfaces. Our observations from the study have revealed extensive use of waypoints [MB97] and annotations by users during the re-finding process. We comment on the implications of these results for designing tools for information re-finding, especially for voice interfaces to re-finding.

## 2. RELATED WORK

Several projects have begun to explore aspects of information re-finding. We summarize these here and note a number of previous projects that have developed tools for information organization and re-use.

### 2.1 Information Re-Finding

Two current projects are closely related to our work and need special mention. Jones et al. [JBD01] and Alvarado, Teevan, Ackerman, and Karger [ATA+03] conducted in-situ studies of people engaging in re-finding tasks. By looking at re-finding in context, these studies have been able to examine what types of re-finding needs people have and how they use existing tools (such as email and bookmarks) to support these re-finding needs.

*Keeping Found Things Found* – Jones et al. have investigated users' behaviors and techniques for organizing and re-accessing information found on the web in their "Keeping Found Things Found" project [JBD01]. They found that people use a variety of methods for re-use and that the choice of method may depend on what function(s) are trying to be supported. For example, sending an email message to oneself with a URL is a good method for re-use if the goal is to support remote access and to serve as a reminding function. Based on their observations, Jones, et al. identified methods and functions that are important for information re-use.

*Haystack* – Alvarado et al. [ATA+03] have investigated several aspects of information re-finding in a study as part of the MIT Haystack project. They conducted an inquiry in which they interrupted 15 participants in their normal work environments twice a day for five days to conduct short, directed interviews regarding their most recent information seeking activities. Among their findings, they identified two main approaches that participants took when looking for information: orienteering and teleporting. Orienteering, as defined in their paper, "involves using contextual information to narrow in on the actual information target, often in a series of steps" and is a type of situated navigation [ATA+03, p.3]. The other approach they observed was teleporting, or an attempt by the user "to take themselves directly to the information they're looking for" [ATA+03, p3]. Teleporting could be viewed as a type of plan-based navigation [JF97].

We have found a number of similar findings between their research and the results we present here, despite very different experimental approaches. For example, we have observed both orienteering and teleporting approaches in our data. Alvarado et al. also made an important observation: "people maintained a large amount of contextual information about the specific piece of information they are looking for" [ATA+03, p.4]. In our study, we also observed that contextual information plays an important role in an iterative re-finding process.

*Waypoints* – The process by which users initially find information may have a significant impact on how they attempt to re-find it [MB97]. In a study of users performing and recalling web searches, Maglio and Barrett [MB97] observed that searchers tended to have routines for searching and that they recalled only a few important sites, or waypoints, on the path to their goal. Participants in their study performed web searches one day and then, on the following day, were asked to "verbally recall" [MB97, p.6] and re-create the searches. Waypoints also appear to play a reminding function. Users may be able to recall certain waypoints, but also may rely on being able to recognize information contained at waypoints to help them get further toward their goal.

*Recall and Recognition* –In a study by Mayes, Draper, McGregor, and Oatley [MDM+88], users of word processing software were not able to recall menu items from the word processor when asked to describe them on a questionnaire. Mayes et al. provided several possible explanations for why users had trouble with recall of the menu items on the questionnaire, but no trouble using them in the word processor. One of their explanations that has relevance here is that if users can rely on information to be found in the environment, they may not commit it to memory because they can rely on re-finding it in the environment when it is needed [MDM+88, p.285]. We believe this is an important concept in how people organize and plan for information re-use and how people approach the process of trying to re-find information.

*Addressability of Information* – Recall and recognition are related to the notion of what we refer to as "the addressability of information" [Ram02, p.12]. Addressability concerns how different paths, connections, access mechanisms and approaches can be used to describe the location of information. For example, for some web sites, a user may recall the specific URL for that site. In this way, the user is addressing the web site directly by its URL. However, some web sites have difficult to remember URLs, or may be accessed infrequently. In these cases, it may be easier to rely on a different form of addressing. For example, a user may rely on knowing that the WWW2004 website can be re-located by going to a web search engine, entering "WWW 2004", and browsing the top results. In this way, the site is accessed in a way that is already familiar to the user (i.e. the search engine is familiar and the search string is familiar) [Ram02].

## 2.2 Tools for Re-Finding

Information found on the web often takes the form of semi-structured data [Abi97] [NAM97]. Several projects at Apple Computer and one at Intel explored the use and manipulation of semi-structured pieces of information, or "information nuggets" [LNW99], contained in larger sources. These projects included Apple Data Detectors [NMW98], LiveDoc [MB98], DropZones [BM98], Grammex [LNW99], and the Intel Selection Recognition Agent [PK97].

Remembrance Agent [RS96], Margin Notes [Rho00], and Haystack [AKS99] attempt to help users collect and use personal stores of information by observing users' interactions with documents and applications. Furthermore, Remembrance Agent, Margin Notes, and Watson [BH00] use information gathered by observing users' interactions with applications to try to make recommendations of other relevant information based on the user's current tasks. A recent article by Steve Lawrence [Law00] provides an excellent survey of systems that make use of contextual information in searching web information.

## 3. WEBCONTEXT

WebContext is a prototype system we have developed as a testbed to explore concepts related to providing a remotely accessible voice interface for re-finding information initially viewed on the Web. We envision WebContext as a system to support the organization and retrieval of information nuggets in an effort to support a wider array of information retrieval for mobile users.

The WebContext architecture is shown in Figure 1. The core of the system is a set of modules for extracting and combining information found on saved web pages. These modules parse HTML web pages and extract pre-defined information nuggets. Initially, we have implemented a handful of routines for extracting information such as phone numbers and addresses. The text of the pages is scanned and keywords and phrases are extracted and indexed to support re-finding. Information extracted from individual web pages is combined using an information combiner module that creates an index that maps information to pages.. The combiner process also generates VoiceXML pages and VoiceXML grammars that allow users to search the index using a voice interface. Additional details of the WebContext architecture and its limitations are given in [CPR01].

Our initial voice interface for WebContext was fairly primitive. The task supported by the voice interface required two pieces of information: words to help identify a web page and the type of information to be retrieved. From a functional point of view, the task was very simple and we built the voice user interface to match the functional representation of the task. A sample dialog with the system is shown in Figure 2. In this sample dialog, a user is trying to find the phone number of a fictitious hotel (the "Anytown hotel"). The user had previously seen a web page for the hotel while browsing the web and remembers the name of the hotel, but not the phone number.

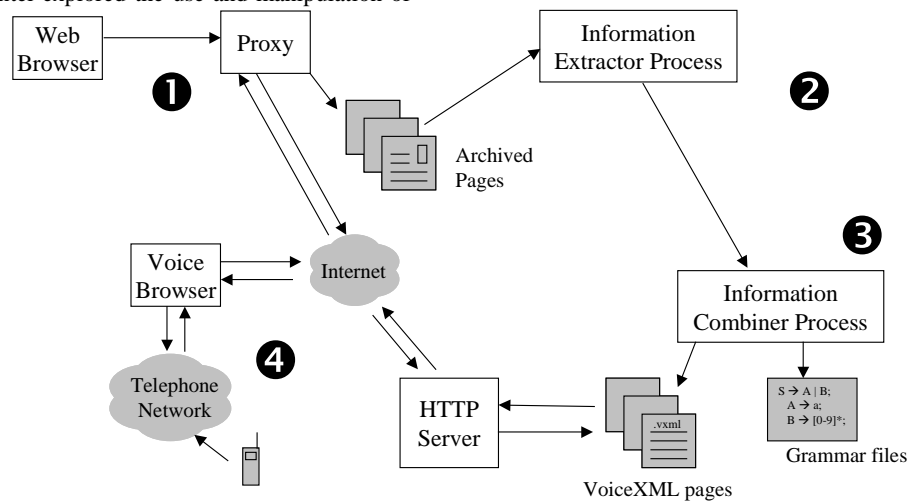

**Figure 1. WebContext Architecture**

In informal evaluations of the initial WebContext interface, we found that users were having difficulty interacting with the system. Our initial approach of having the system drive the process through an up-front series of questions regarding keywords on the page and what specific information the user was looking for did not work well for all users in our informal testing. One user thought that the interface approached the re-finding task in an opposite fashion to how she wanted to approach re-finding. We began exploring ideas about how to improve the system as a collaborative partner in re-finding. The initial system did not expose the information extraction and storage capabilities of the system, so users may have had difficulty constructing a mental model about what the system was doing and how to converse with it. We set out to fix this as part of an effort to increase the shared context between the user and the system.

One idea was to allow users to explicitly annotate information to be saved. Web annotation systems allow users to annotate information on web pages using various annotation mechanisms. These systems provide a possible way for users to designate information and web pages that contained important information.

Ultimately, we decided that the best way to explore these ideas for a more collaborative interface was to conduct an empirical study to examine how users approach re-finding, how they express information re-finding requests, and how annotations might facilitate re-finding tasks.

| [1] | User: | <calls into the WebContext system> |
|---|---|---|
| [2] | WC: | Welcome to the WebContext system. Please say some words to help identify the pages to search. |
| [3] | User: | Anytown hotel |
| [4] | WC: | What piece of information are you looking for? |
| [5] | User: | The phone number. |
| [6] | WC: | Looking for phone numbers on web pages with Anytown hotel. Is this correct? |
| [7] | User: | Yes. |
| [8] | WC: | Now looking for matches. {pause}On the page titled, "Anytown Hotel Home Page," there is one result, {pause}five five five {pause} one two three four. |

**Figure 2. Example Dialog with the Initial WebContext Voice Interface**

## 4. Re-Finding Study

In this section, we describe the study we conducted to explore the approach users take to re-finding information. We had three goals for the study. First, we were interested in observing the fluid process that users follow to re-find previously seen information. To explore this, our study involved collaborative dialogues between two participants to observe the step-by-step process followed in re-finding. Second, we wanted to study the effect that information artifacts have in the re-finding process. Finally, we were interested in exploring if users could create information artifacts that could later be used in the re-finding process.

Three key features distinguish our study from previous research on re-finding:

Controlled, Laboratory Study – The study was conducted as a controlled, laboratory study rather than a contextual inquiry. This allowed us to examine re-finding behaviors across identical tasks and similar situations.

Collaborative Dialog – We structured the study as a collaborative dialog over a telephone between two participants, one who had access to the information and one who was trying to re-find the information. This protocol was chosen because it provides good insight into the re-finding process (possibly better than talk-aloud with one participant), and also so that we could examine characteristics of the dialogs for evidence of how contextual information and shared context was used in the re-finding process. It also was chosen to support our interest in voice interfaces for information retrieval.

Explicit Context – We introduced the ability for participants to make and save annotations on web pages as they found information. This was done to explore if and how explicitly generated contextual information would be used in the re-finding process.

### 4.1 Participants
A total of 12 participants in six groups of two participated in the study. Participants were recruited from the Virginia Tech community and a majority were graduate students in Computer Science or Human Factors. Because this study examined dialog and language use, participants were required to speak North American English as their native language. All participants were familiar with web browsing.

### 4.2 Sessions and Tasks
The study consisted of two sessions that each lasted approximately one hour. In the first session, a participant (who we will refer to as the User) completed a set of tasks that involved finding information on the Internet using a web browser. The second session was scheduled about a week later and involved both the User from the first session, and a second participant (who we will refer to as the Retriever). In the second session, the Retriever helped the User complete tasks that involved re-finding information the User had found during the first session. Recordings were made of the sessions and the interactions between the participants. Additional details of each session are given below.

First Session – The first session involved only the User. In this session, the User was given a set of five tasks that involved finding information on the Internet using a web browser. The five tasks were, in order, 1) finding two movie showtimes at two theaters for three movies, 2) finding the phone numbers and addresses for four nearby restaurants, 3) finding information (event name, location, price, and hours) about three events or tourist locations for a trip to San Francisco, 4) finding names, price ranges, and phone numbers for restaurants in San Francisco for four different types of cuisine (Italian, Chinese, Thai, and American), 5) a user-defined task that allowed the user to decide a specific piece of information to look up on the web. These tasks were selected to provide a variety of directed and freeform information finding tasks.

The web browser was equipped with commercial software that allowed the User to make web annotations (such as highlighting, drawing, and notes) on web pages. Annotations became associated with that page so that whenever it was re-accessed, the annotations were re-displayed also. Each annotation could be given a classification. Three classification categories were made available by default: movies, restaurants, and travel. Users were also able to create their own classifications categories.

Prior to beginning the first session, Users were shown a video of instructions that explained the tasks and interactions that would take place in both the first and second sessions. This video described the role of both the User and the Retriever, and was shown to all participants. Participants were also given training on how to use the web annotation tools.

Users were instructed, 1) that they could make as many or as few annotations and classifications on web pages as they wished, 2) that all the web pages they browsed were being saved in a history log, and 3) that the retriever would have access to all their annotations and history during the second session to help them re-find information.

Users were given 45 minutes to work on the five tasks. Each task instruction page included a place for users to write down their findings as they completed the task. After 45 minutes, if the user had not completed the tasks, the experimenter notified the User and gave them the option of finishing the current task, up to a limit of one hour total.

Second Session – The second session was scheduled approximately one week after the first session and involved both the User and the Retriever. The Retriever was asked to arrive first and was shown the instruction video and given training on how to access the annotations and history log. The web annotation tools supported the retrieval of pages and annotations. Listings of pages with annotations could be viewed in a sidebar of the web browser and could be organized by web site and by classification label.

When the User arrived, they were seated in a different room from the Retriever. The User was presented with a new set of tasks that involved re-finding information that had been found during the first session. The User was given the same number of tasks as they completed in the first session; this was either four or five tasks for all participants. The re-finding tasks mirrored the finding tasks that had been given during the first session. The five tasks were, in order, 1) remember or re-find the name of a movie and re-find the earliest showtimes at two theaters, also re-find the rating for the movie, 2) remember or re-find the names of two restaurants and re-find their phone numbers, 3) re-find the names and locations of all the events or tourist activities related to San Francisco that were found during the first session, 4) re-find the names and addresses of one Italian and one Chinese restaurant in San Francisco, 5) try to re-find the information from the user-defined task from the first session. These tasks were selected so that they mirrored the finding tasks from the first session, but provided some variety in the information requested. In some cases, these tasks requested that users re-find a subset of information found in the counterpart task from the first session. In other cases, the task required re-finding the same path, but requested different (new) specific information. For example, the movie re-finding task (task 1) asked Users to find the movie rating although the rating was not asked for in the first session movie task.

The User did not have access to any of their information from the first session, but the Retriever did. The Retriever was seated at the computer that the User had used during the first session and had access to a complete history of the web pages that the User viewed on the first day. The Retriever also had access to any web annotations and classifications made by the User as they searched.

In the instructions, Users were informed that they should direct the re-finding process and not to simply "off-load" the task on the Retriever. Users placed telephone calls to the Retriever to accomplish the re-finding tasks.

## 5. ANALYSIS

We report here on our analysis of tasks 1-4 for all pairs of participants. Several Users did not complete the fifth task, so we have excluded it from this analysis. A total of 26 separate telephone conversations were collected for the six user-retriever pairs for tasks 1-4. Twenty-six conversations were collected instead of 24 because there were two instances in which participants made two phone calls as part of one task. In one case, the re-finding task description allowed users to break up the task into two parts if desired and one user did so. In the second case, one pair of participants was unable to complete the task on the first try, asked the experimenter a clarifying question, and decided to try the task again.

Transcriptions were made of the 26 conversations between the Users and Retrievers. These transcriptions were verified and then coded for conversational phases, instances of common ground, use of waypoints, use of annotations, specific information requests, and additional recalled and recognized items. Coding was conducted in three stages. In the first stage, one of the authors of the paper developed an initial coding scheme and completely coded the data. In a second stage, the coding scheme was explained to a second coder who then completely coded the data. Then, in a final stage, the two coders jointly coded the data a third time, reconciling their individual coding and making small adjustments to the coding scheme.

## 6. RESULTS AND OBSERVATIONS

In this section, we present results and observations from our study. We describe two findings here: 1) the re-finding process relied heavily on the use of contextual information and domain artifacts to move closer to the goal; and 2) explicitly added artifacts (i.e. annotations) can be used to expand the addressability of information and generate additional shared context.

After completing our study and considering the results of Alvarado et al [ATA+03], we can summarize our current view of re-finding in a simple observation: re-finding often relies on using contextual information in an iterative process.

## 6.1 Reliance on Artifacts and Context

Users and retrievers relied on a number of artifacts and contextual information as part of making progress in the iterative re-finding process. Specifically, waypoints and annotations were used to help achieve points of grounding where the User and Retriever reached a common understanding of a web page, annotation, goal, or piece of information. In this section, we will present results regarding the use of two such artifacts: waypoints and annotations.

*Waypoints*

We observed extensive use of waypoints [MB97] by users in their attempts to re-locate sources of information. In our analysis involving waypoint usage, we have adopted a less restrictive view of waypoints than Magilo and Barrett [MB97]; we have dropped their requirement that the waypoint definitively be along the path to the goal, and instead focus on any mention of a specific node. This allows us to consider as waypoints even nodes that Users may mis-remember as being on the path to the goal (it is a waypoint to the User).

Waypoints were used in 20 of the 26 conversations (76.9%) we observed. The average number of waypoints per conversation was 3.46 (stdev = 4.26). Some participant pairs made extensive use of waypoints in their re-finding while other pairs made less use of them. In some cases, this was due to reliance on other artifacts such as annotations and descriptions of the information being sought. However, in some dialogues, the User and Retriever managed to achieve goals without much use of either waypoints or annotations. This was especially true for one particular pair of participants. In many of their re-finding dialogues, no waypoints or annotations were used, but the Retriever was especially adept at locating and finding the information requested.

To investigate how waypoints were used, we classified waypoints into three main categories: Page/Site Titles, URLs, and Page Descriptions. Each of these is described below.

Page and Site Titles – This refers to full and partial names of web sites and web pages and also to names of groups or entities associated with pages and sites. Some examples from our data include: "the Outback Steakhouse website," and "Regal Cinema site."

URLs – These are spoken references to URLs and were often formed as "<name> dot com", to refer to the top-level "home" page for a particular web site. In many cases, these references communicate both a URL and a site title. For example, we observed, "Fandango dot com," "W W W dot Macados dot com," and "Movie of Yahoo dot com."

Page Descriptions – This refers to descriptions of the contents of a web page or site. Some examples from our data include: "it's kind of like a Yellow Pages kind of thing," and "a sort of general page listing {pause} of many different restaurants in Blacksburg and their addresses and phone numbers and such."

Both Users and Retrievers provided page descriptions, and sometimes the description became a collaborative process that helped solidify that they had reached common ground at a page. The example shown in Figure 2 illustrates a case in which the Retriever completed a page description for the User (turn 2). Based on the common ground of the page, the user then quickly made a request for specific information believed to be on the page.

Figure 3 shows the usage of the three categories of waypoints (URLs, Titles, Descriptions) for Users and Retrievers across all dialogues.

As can be seen in Figure 3, both Users and Retrievers made use of waypoints to help the re-finding process. However, Users made more references to specific URLs. URLs can be used in attempts to teleport to information, so it is not surprising to see more use of them by Users than Retrievers. Another note to be made here concerns recall versus recognition. In many cases where Users made use of a waypoint, it was something they recalled and used to help navigate to a source of information. Retrievers often presented waypoints to Users for recognition in order to help the navigation process.

| [1] | U: | Yeah, it should, like, have, like a lots of stuff up at the top, but the movies are actually, like, down… |
| [2] | R: | In front of the page. |
| [3] | U: | Yeah. |
| [4] | R: | I see a little {unintelligible} of Men In Black Two. |
| [5] | U: | And, {unintelligible} again, the earliest times. {pause}What's the name of that theater anyways? |

**Figure 2. Example of Collaborative Grounding on a Page Description**

|       | Users | | Retrievers | |
|-------|-------|-------|-------|-------|
|       | # | %User | # | %Ret |
| URLs  | 8 | 17.8% | 1 | 1.8% |
| Titles | 17 | 37.8% | 27 | 48.2% |
| Desc. | 20 | 44.4% | 28 | 50.0% |
| Total | 45 | 100.0% | 56 | 100.0% |

**Figure 3. Waypoint Usage by Type**

Next, we examine the use of another type of artifact used to help re-find information, annotations.

*Annotations*

As with waypoints, annotations were a type of artifact used to help re-locate sources of information. However, annotations differ from waypoints in that we added the ability for users to explicitly create the annotations, while waypoints were an implicit piece of context that are a naturally occurring aspect of web browsing. We introduced the ability for users to create annotations in order to examine if and how they would be used to help re-finding.

So that Users would understand the possible value of annotations, Users were told in the first session that the annotations they made would be available to the Retrievers during the second session. However, Users were told that they were free to make as many or as few annotations as they wished.

Annotations were referenced in 22 of the 26 conversations (84.6%) we collected. The average number of annotations per conversation was 6.83 (stdev = 7.38). To investigate how annotations were used, we classified annotation references into three main categories: Category Names (Cat), Annotation Type (Type), and general references to annotations (Ref). Each of these will be described below.

Category Names. Each annotation could be associated with a named category. Examples of references to category names from our data include: 1) "User: So there should be actually a section, in my annotations, called restaurants", and 2) "User: And I believe if you go into my notations… If you click on the matinee one… it'll pull up some stuff…"

Annotation Types. There were several types of annotations that could be made on the web pages: text could be highlighted, text notes could be added, and drawings could be made on the pages (such as circling items or putting arrows or an "X" next to them). Users and Retrievers made references to these specific types of annotations.

References to Annotations. Sometimes a reference would be made not to a specific annotation type or category name, but just to the annotation feature in general. For example, "Retriever: Do you remember how you annotated it?"

Figure 4 shows the usage of the three classes of annotations (Categories, Types, and References) for Users and Retrievers across all dialogues.

|       | Users |        | Retrievers |       |
|-------|-------|--------|------------|-------|
|       | #     | %User  | #          | %Ret  |
| Cat   | 28    | 35.0%  | 28         | 31.5% |
| Type  | 20    | 25.0%  | 45         | 50.6% |
| Ref   | 32    | 40.0%  | 16         | 18.0% |
| Total | 80    | 100.0% | 89         | 100.0%|

**Figure 4. Annotation Usage by Type**

Users often included references to annotations (Ref) as part of navigation suggestions (i.e. "if you go into my notations") and this may explain the higher percentage of references (Ref) by Users than Retrievers in Figure 4. Similarly, Retrievers often provided descriptions that included the annotation types they were looking at on the screen (i.e. "those are the only two circled") and this may account for their higher percentage of references to annotation types (Type).

## 6.2 Artifacts to Expand Addressability

Allowing Users to make annotations on the web pages was a feature that was included in our study to allow us to examine if and how explicitly added contextual information would be used in the re-finding process. Our results (as summarized in Figure 4) indicate that both Users and Retrievers did make use of annotations and that the total number of annotation references across all conversations (80+89=169) was slightly higher than the number of waypoints used (45 + 56 = 101). In addition, Users made 35% of their annotation references to categories, indicating that they had good recall of the locations of their annotations.

This data shows a that an artifact explicitly added by the User can be useful in the re-finding process and can help increase the addressability of information sources and increase shared context.

## 7. DISCUSSION

Based on our observations, users often make use of contextual information and domain artifacts such as waypoints to facilitate the search process. Our results are consistent and support both Maglio and Barrett's findings on the use of waypoints [MB97] and Alavarado et al.'s findings regarding the orienteering approaches to re-finding they observed [ATA+03]. Our results are also consistent with concepts from the navigation of electronic spaces [JF97]. We contribute that reaching points of common ground appear to play an important role in the re-finding process.

Another characteristic observed is that users take an iterative approach to re-finding information. Future work should explore if a search tool for re-finding should support this approach of making incremental progress towards the goal. Research could also examine how search tools can address user needs for re-finding. Alvarado et al. have suggested that users may prefer the orienteering approach over a search-engine-style keyword search [ATA+03]. It is possible that information found on the web may not be remembered by the addressability of the information, that is, not by the path needed to reach the information, but instead by a process of how to reach the information. Research on waypoints [MB97], information scent [CPC+01], recall/recognition [MDM+88], orienteering/teleporting [ATA+03], and our work seem to support this incremental process of information re-finding.

Different artifacts are used to guide the waypoint identification process. The artifacts used in our study included URLs, page titles, and general descriptions of page content. We also found that new artifacts can be added by the user and used successfully to help re-finding. In our study, these new artifacts were in form of annotations. When they are added, we found that they could be used to help in the re-finding process and increase the addressability of the information being sought. These extra cues become integrated as parts of the context that the user looks for when trying to re-find information.

From a conversational point of view, we can say that the annotations are part of the *shared context* between the User and the information that was already seen on the web. This has a particular significance; since the annotation was explicitly added, it is one of the only artifacts that a re-finding tool can be certain is part of the shared context with the user. The user might not have seen the URL, title, or even parts of a page, but since the user explicitly added the annotation, there is more certainty that the user was at some point aware of the annotation. However, creating annotations requires user effort. Future work would need to explore the willingness of users to invest this time versus the potential benefit in re-finding.

With the current proliferation of mobile devices (cell phones, PDAs, MP3 players, etc.), the availability of information access from these devices, and users' needs to re-access information while away from their primary computer, it is important to explore information re-finding from a mobile standpoint. To this end, we need to understand the re-finding process that users follow well enough to support it from different tools and devices. In the mobile work environment that is available today, users need to re-find information, and often from very different devices. The study described here begins to shed light on the interactive process that users follow to re-find information and the artifacts used to achieve re-finding. Our study was conducted using a protocol that separated the User during their re-finding tasks from the computer with which they first found the information. Our results show the possibility of using annotations as context for increased information addressability and possibilities for making them available remotely from a different interaction medium.

## 8. ACKNOWLEDGMENTS
We thank Miranda Capra for her support and assistance with data analysis, and Dr. Naren Ramakrishnan for his continued support and feedback. This work was supported in part by the National Science Foundation under Grant No. IIS-0049075, and by a grant from IBM to explore the use of VoiceXML within their WebSphere product.